\documentclass[aps,prl]{revtex4}
\usepackage{amsmath,amssymb}
\usepackage{graphicx}
\usepackage{dcolumn}
\usepackage{bm}
\usepackage{color}

\begin{document}

\title{Chiral filtration of light by Weyl-semimetal medium}
\author{ Nikolay M. Chtchelkatchev$^{1,2,3}\thanks{%
E-mail contact:shchelkachev.nm@mipt.ru}$, Oleg L. Berman$^{4,5}$,
Roman Ya. Kezerashvili$^{4,5}$, and Yurii E. Lozovik$^{6,7}$ }
\affiliation{$^{1}$Vereshchagin Institute for High Pressure Physics, Russian Academy of Sciences, 108840 Troitsk, Moscow, Russia\\
$^{2}$Moscow Institute of Physics and Technology, 141700 Dolgoprudny, Moscow Region, Russia\\
$^{3}$Ural Federal University 620002, Ekaterinburg, Russia\\
$^{4}$New York City College of Technology, The City University of New York, Brooklyn, NY 11201, USA\\
$^{5}$The Graduate School and University Center, The City University of New
York, New York, NY 10016, USA\\
$^{6}$Institute for Spectroscopy, Russian Academy of Sciences, Troitsk,
Moscow,108840 Russia \\
$^{7}$MIEM at National Research University Higher School of Economics,
Moscow, Russia
}
\date{\today}

\begin{abstract}
Recently discussed topological materials  Weyl-semimetals (WSs)
combine both: high electron mobility comparable with graphene and
unique topological protection of Dirac points. We present novel
results related to electromagnetic field propagation through WSs. It
is predicted that transmission of the normally incident polarized
electromagnetic wave (EMW) through the WS strongly depends on the
orientation of polarization with respect to a gyration vector
$\mathbf{g}$. The latter is related to the vector-parameter
$\mathbf{b}$, which represents the separation between the Weyl nodes
of opposite chirality in the first Brillouin zone. By changing the
polarization of the incident EMW with respect to the gyration vector
$\mathbf{g}$ the system undergoes the transition from the isotropic
dielectric to the medium with Kerr- or Faraday-like rotation of
polarization and finally to the system with chiral selective
electromagnetic field. It is shown that WSs can be applied as the
polarization filters.
\end{abstract}

\maketitle

%\pacs{PACS}

%\section{Introduction}

The experimental discovery of the 3D Dirac fermions in $\mathrm{Na_{3}Bi}$
and $\mathrm{Cd_{3}As_{2}}$~\cite{Liu_14,Borisenko,Neupane} has opened a new
window of possibilities in condensed matter and material science with Weyl
semimetals (WSs)~\cite{Xiong,Wang,Liang,Galletti,Zhang,Puphal}. These materials are characterized by broken time-reversal
or inversion symmetry. The Brillouin zone of a WS contains pairs of Weyl
nodes. Since these nodes can become sources and sinks of Berry curvature,
unusual surface states can be formed~\cite{Wan}. WS can be viewed as a 3D
generalization of graphene, where the Dirac points are not gapped by the
spin-orbit interaction, and the crossing of the linear dispersions is
protected by crystal symmetry~\cite{Young}. The Dirac nature of the
quasiparticles in WS was confirmed by investigating the electronic structure
of these materials with angle-resolved photoemission spectroscopy and a very
high electron mobility was observed, up to $2.8 \times 10^{5} \ \mathrm{cm^{2}/Vs}$~%
\cite{Liu_14,Borisenko,Neupane}, which is comparable to that of the
best graphene. The magnetotransport in Dirac materials was studied
in Ref.~\cite{Abanov}. The optical conductivity tensor of a 3D Weyl
semimetal was evaluated from semiclassical Boltzmann transport
theory~\cite{Polini}. Nontrivial topology of WS manifests itself in
the interesting optical properties~\cite{Kotov,Belyanin}. A new way
to isolate real topological signatures of bulk states in WSs was
developed~\cite{Folkes}. Properties of WSs are reviewed in
Ref.~\cite{Yan,Armitage_rmp}.

In this letter we investigate the interplay between the topological nature of {WSs and their%
} optical properties. It is predicted that WS {can} behave as broadband chiral
optical medium that selectively transmits and reflects circularly polarized
electromagnetic field in wide range of frequencies. The search for a chiral
optical medium has been marked by the discovery of liquid crystals having
chiral selective transmission and reflection~\cite{Belyakov}. Later photonic
crystals and metamaterials with similar properties, but in a narrower
frequency range, have been manufactured~\cite{Vetrov}. In this paper we predict that
topologically nontrivial solid state crystals - Weyl semimetals may behave
like a chiral optical medium.

In WSs the valence and conduction bands touch each other at the isolated
points of the Brillouin zone, termed as Weyl nodes, close to the energy
level of the chemical potential $\mu $. The Hamiltonian $\hat{H}$ of the
charge carriers in the momentum space near a Weyl node is given by $\hat{H}%
=\chi v_{F}\mathbf{p}\cdot \hat{\sigma}-\mu $ \cite{Vafek,Wehling}, where $%
\chi =\pm 1$ is the chirality, $v_{F}$ is the Fermi velocity, $\mathbf{p}$
is the momentum, and $\hat{\sigma}$ represents Pauli matrices. A doped Weyl
semimetal with a positive chemical potential $\mu >0$ is under consideration
in this paper.

Weyl nodes of WS form pairs of opposite chirality~\cite%
{Nielsen1,Nielsen2,Nielsen3}. Let us consider a WS with two Weyl nodes,
which are separated in the first Brillouin zone by the wave vector $\mathbf{b%
}$. Then the electromagnetic term of WS action acquires the topological $%
\theta $-term $S_{\theta }=\frac{e^{2}}{4\pi \hbar }\int dt\int
d^{3}r\,\theta \mathbf{E}\cdot \mathbf{B}$ \cite{japan,Burkov,Tewari,Hosur}.
In the latter expression $e$ is the electron charge, $\mathbf{r}$ is the
coordinate vector, $t$ is time, $\chi =\pm 1$ and $\theta =2\left( \mathbf{b}%
\cdot \mathbf{r}-b_{0}t\right) $. If $\theta =0$ the system is not
characterized by the topological properties.

The $\theta $-term in the action results in the appearance of two nontrivial
terms in the relation between the displacement field $\mathbf{D}$ and the
electric field $\mathbf{E}$ in WS~\cite%
{Wilczek,Grushin,Hosur2,Trivedi,Zyuzin,Polini,Sharma,DasSarma,Rosenstein}:
%-------------------------------------------------------------------------
\begin{eqnarray}  \label{defr}
\mathbf{D}=\varepsilon _{0}(\omega )\mathbf{E}+\frac{e^{2}}{\pi
\hbar \omega
}(\nabla \theta )\times \mathbf{E}+\frac{ie^{2}}{\pi \hbar c\omega }\dot{%
\theta}\mathbf{B},
\end{eqnarray}
%-------------------------------------------------------------------------
where the last two term present the anisotropy in the dielectric
tensor. The the isotropic term $\varepsilon _{0}(\omega )$ in the
dielectric tensor is related to the Coulomb interaction between
electrons. At zero temperature in the framework of a tight-binding
model with a strong spin-orbit interaction in Ref.~\cite{Rosenstein}
the complex dielectric constant for 3D semimetals was calculated as
$\varepsilon _{0}(\omega )=\varepsilon ^{\prime }(\omega
)+i\varepsilon ^{\prime \prime }(\omega )$ with

\begin{eqnarray}  \label{diel}
\varepsilon ^{\prime }(\omega ) &=&1+A\ln \left( \frac{B}{\omega }\right) %
\left[ 1+F\ln \left( \frac{B}{\omega }\right) \right] ,  \notag \\
\varepsilon ^{\prime \prime }(\omega ) &=&D\left[ 1+\alpha \left( E\ln
\left( \frac{\omega }{\varpi }\right) +C\right) \right],
\end{eqnarray}
when the ultraviolet cutoff frequency $\bar{\omega}${$\gg {\omega }.$ In
Eqs. \eqref{diel} $A=\frac{N_{W}e^{2}}{3\pi v_{F}\hbar}$, where
$N_{W}=2$ is the number of Weyl points, $v_{F}=10^{6}$ m/s, $B=\Lambda v_{F}$, where $\Lambda =%
\frac{\pi }{a_{0}}$ is the UV cutoff and $a_{0}=3$ ${\AA}$ is the
length of a lattice vector, $C=-\frac{5}{3\pi }$, $D=$ $\frac{N_{W}e^{2}}{%
6v_{F}\hbar}$, $E=\frac{2}{3\pi }$, $F=\frac{\alpha }{3\pi },$ and $\alpha =%
\frac{e^{2}}{\epsilon _{W} v_{F}\hbar}$, where $\epsilon _{W}=3$ is the intrinsic
dielectric constant.}

\begin{figure}[tbp]
\centering
\includegraphics[width=8.5cm]{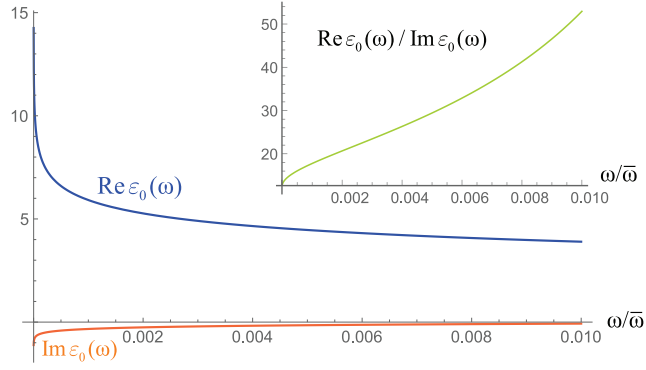}
\caption{(Color online) For chosen parameters, $\protect\varepsilon %
_{0}^{\prime }(\protect\omega )\gg |\protect\varepsilon _{0}^{\prime \prime
}(\protect\omega )|$. Thus, the attenuation of electromagnetic waves due to $%
|\protect\varepsilon _{0}^{\prime \prime }|\neq 0$ is weak.}
\label{figReImeps}
\end{figure}

\begin{figure}[tbp]
\centering
\includegraphics[width=8.5cm]{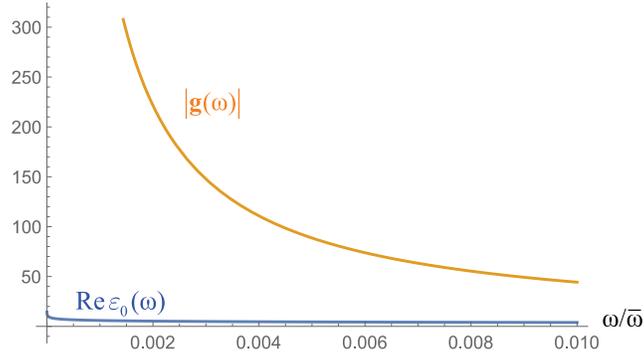}
\caption{(Color online) Comparison of the first and second terms of the
dielectric tensor of a WS. }
\label{figg}
\end{figure}
\begin{figure}[tbp]
\centering
\includegraphics[width=8.5cm]{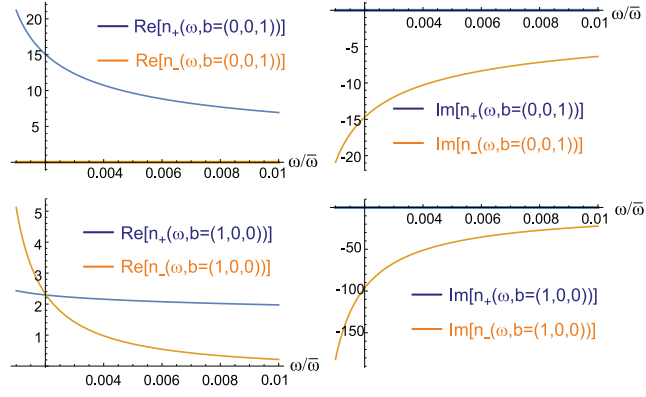}
\caption{(Color online) Refraction coefficients at different orientations of
$\mathbf{b}$. Since $|\mathrm{\ Im}N_{-}|\gg 1$, electromagnetic waves in $z$%
-direction with \textquotedblleft -\textquotedblright\ chirality can not
propagate.}
\label{fig:Npm}
\end{figure}

The ultraviolet cutoff frequency $\bar{\omega}$ can be defined from the
following expression for the density of electronic states, which equals the
atomic concentration $n$ in the tight-binding model:
%%%%%%%%%%%%%%%%%%%%%%%%%%%%%%%%%%%%%%%%%%%%%%%%%%%%%%%%%%%%%%%%%%%%%%%%%%%%%%%%%%%%%%%%%%%%%%%%%%%%%%%%%%%%%%%%%%%%%%%%%%%%%%%%%%%%%%%%%%%%%%%%%%%%%%%%%%%
\begin{eqnarray}
n=2\int \frac{d^{3}p}{(2\pi \hbar )^{3}}=\frac{1}{\pi ^{2}\hbar ^{3}}%
\int_{0}^{\hbar \bar{\omega}}\frac{\mathcal{E}^{2}}{v_{F}^{3}}d\mathcal{E}=%
\frac{\left( \bar{\omega}\right) ^{3}}{3\pi ^{2}v_{F}^{3}}\ ,
\end{eqnarray}%
%
%%%%%%%%%%%%%%%%%%%%%%%%%%%%%%%%%%%%%%%%%%%%%%%%%%%%%%%%%%%%%%%%%%%%%%%%%%%%%%%%%%%%%%%%%%%%%%%%%%%%%%%%%%%%%%%%%%%%%%%%%%%%%%%%%%%%%%%%%%%%%%%%%%%%%%%%%%%
where the factor of $2$ in front of the integral stands for two valleys, and{%
\ $\mathcal{E}=v_{F}p$} is the Dirac energy spectrum of the electrons in
WSs. In our calculations we scaled the frequency in units of $\bar{\omega}=%
\sqrt[3]{3\pi ^{2}nc}v_{F}$ as the energy (frequency) unit in all figures.
For material parameters given above, $\bar{\omega}\approx 2\pi \times
1.6\times 10^{15}$~Hz.

In fact for given parameters $\varepsilon _{0}^{\prime }(\omega )\gg
|\varepsilon _{0}^{\prime \prime }(\omega )|$, as it is seen in Fig.~\ref%
{figReImeps}. Thus, the attenuation of electromagnetic waves (EMWs) due to $%
\varepsilon _{0}^{\prime \prime }\neq 0$ is weak.

As it follows from Eq.~\eqref{defr} (see Supplementary material~A, the
dielectric tensor of a WS can be written as
\begin{equation}
\varepsilon _{\alpha \beta }(\omega )=\varepsilon _{0}(\omega )\delta
_{\alpha \beta }+i\epsilon _{\alpha \beta \gamma }g_{\gamma }(\omega ).
\label{eq:eps}
\end{equation}%
In Eq. \eqref{eq:eps} $\epsilon _{\alpha \beta \gamma }$, where $\alpha
\beta \gamma $ corresponds $xyz,$ is the fully antisymmetric tensor, $%
g_{\gamma }=-a(\omega )b_{\gamma }$ is the gyration vector~\cite%
{landau2013electrodynamics} and $a(\omega )=\frac{2e^{2}}{\pi \hbar \omega }$
is an odd real function of $\omega $, $a(\omega )=-a(-\omega )$. The second
term in the dielectric tensor \eqref{eq:eps} is the{\ Hermitian} matrix and
so it does not produce any dissipation despite the fact that it is purely
imaginary.

The dielectric tensor ~\eqref{eq:eps} is well known in magnetooptics~\cite%
{landau2013electrodynamics}, where the gyration vector $\mathbf{g}$ is
typically $\omega $ independent and it is generally small compared to $%
\varepsilon _{0}$ (to first order, $\mathbf{g}$ is proportional to
the applied magnetic field, it is small, and it is considered as a
\textquotedblleft relativistic effect\textquotedblright ). In
contrast, for a WS the situation is opposite: $|\mathbf{g}|\sim
|\varepsilon _{0}|$ (see Fig.~\ref{figg}), $\mathbf{g}$ is
essentially $\omega $-dependent and the direction of $\mathbf{g}$ is
defined by the \textquotedblleft built in\textquotedblright\ into WS
material vector $\mathbf{b}$ related to the Weil points.

\begin{figure*}[tbp]
\centering
\includegraphics[width=\textwidth]{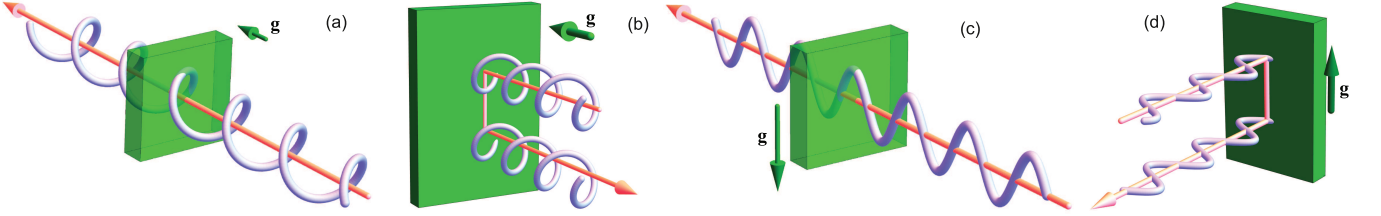}
\caption{(Color online) The figure schematically shows the filtering of
light by polarization by the Weyl semimetal depending on the direction of
the gyration vector $\mathbf{g}$ (shown by green arrows). In (a) and (b) $%
\mathbf{g}$ and $\mathbf{k}$ are parallel. Then EMWs (a) transmit
through WS for one chirality (type of circular polarization); (b)
reflect for the other chirality (opposite circular polarization). In
(c) and (d) $\mathbf{g}$ and $\mathbf{k}$ are perpendicular. Then
EMWs (c) transmit for collinear with $\mathbf{g}$ linear
polarization and (d) reflect otherwise.} \label{figsp}
\end{figure*}
\begin{figure}[tbp]
\centering
\includegraphics[width=8.5cm]{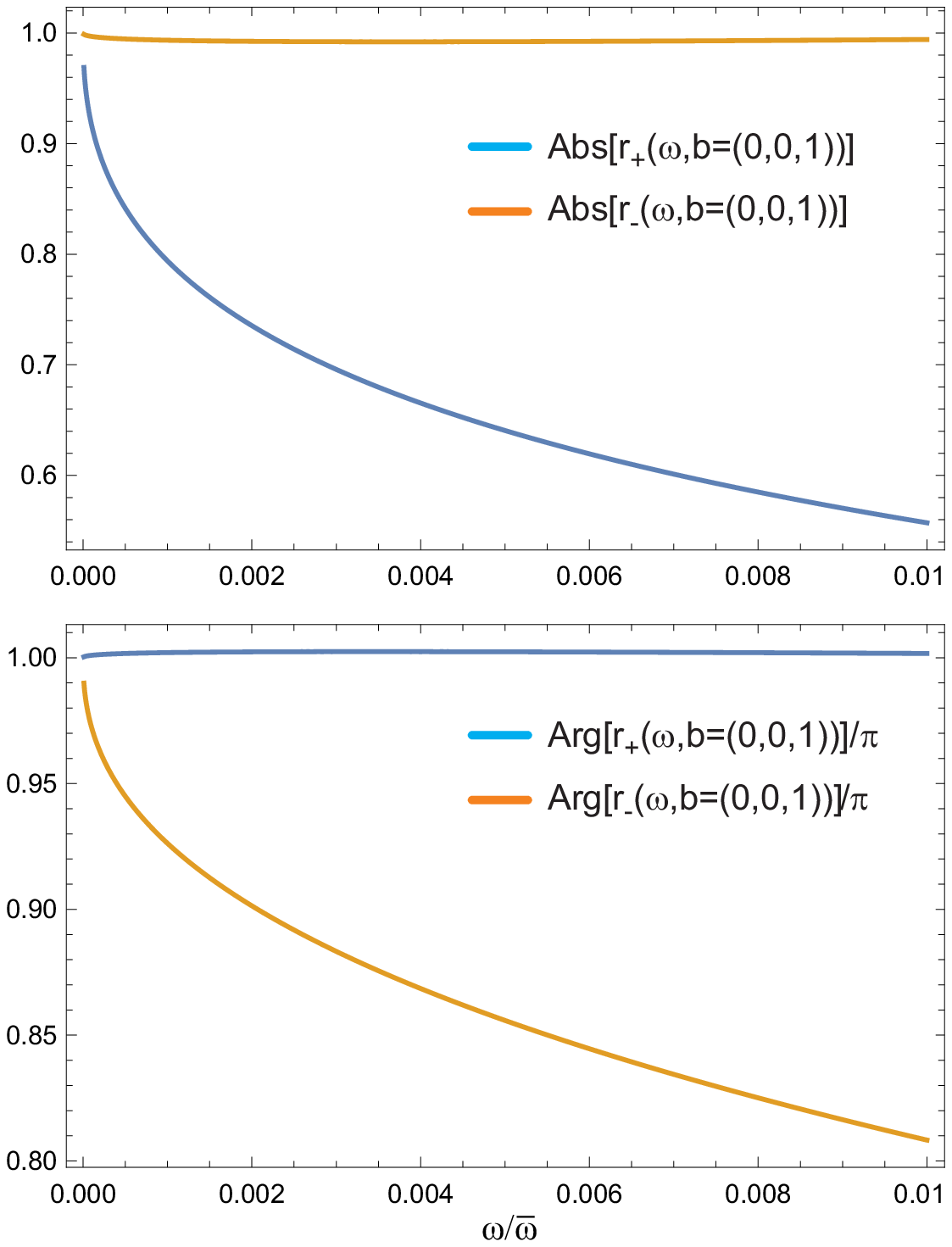}
\caption{(Color online) Reflection coefficients from the Weyl-semimetal
half-space ($z>0$) when $b=(0,0,1)$.}
\label{fig_r_bz}
\end{figure}

We can estimate the amplitude of the second antisymmetric term in the
dielectric tensor: $g(\omega)=a(\omega) b=\frac{2e^2b}{\pi\hbar \omega}=%
\frac{e^2}{\hbar c}\frac{2c\Lambda}{\pi\bar\omega}(b/\Lambda)\frac{\bar
\omega}\omega\approx 1.45 (b/\Lambda)\frac{\bar \omega}\omega$. Typically $%
\omega\ll\bar\omega$ while $b/\Lambda\sim 1$, so $a(\omega) b\gg 1$. This is
illustrated in Fig.~\ref{figg}.

Note that we neglect  surface induced contribution to the reflection
coefficient because the contribution to the reflection coefficient
from the surface is much less than the contribution from the
bulk~\cite{Belyanin}.

We consider a monochromatic EMW in the WS medium with the dielectric
tensor~\eqref{eq:eps}. Then for the electric field {\ }
\begin{equation}
\mathbf{E}=A\mathbf{p}\exp (i\mathbf{k}\cdot \mathbf{r}-i\omega t),
\end{equation}%
where $A$ is the amplitude, $\mathbf{p}$ is the unit vector fixing
the wave polarization, we introduce the vector
$\mathbf{n}=c\mathbf{k}/\omega $. The absolute value of
$n=|\mathbf{n}|$ is usually treated as the refraction coefficient
for the wave. From the Maxwell equations it follows that
\begin{equation}
(n^{2}\delta _{ik}-n_{i}n_{k}-\varepsilon _{ik})E_{k}=0.  \label{eq:Fresnel}
\end{equation}%
This condition known as Fresnel equation allows to find $\mathbf{p}$ and{\ $%
\mathbf{n}$} for a given direction of the wave. Assuming that the
EMW goes along the $z$-direction and using expressions for $%
\varepsilon _{ik}$ from Supplementary material A (see Eq.~(S4)),
one can conclude that Eq.~\eqref{eq:Fresnel} has non-trivial
solutions only if the following condition holds:
%-------------------------------------------------------------------------
\begin{eqnarray}  \label{deter}
\mathrm{det}\left\vert
\begin{array}{ccc}
n^{2}-\varepsilon _{0} & ig_{z} & -ig_{y} \\
-ig_{z} & n^{2}-\varepsilon _{0} & ig_{x} \\
ig_{y} & -ig_{x} & -\varepsilon _{0}%
\end{array}%
\right\vert =0.
\end{eqnarray}
%
%-------------------------------------------------------------------------
Eq.~(\ref{deter}) can be reduced to the quadratic equation for
$n^{2}$,
presented in Supplementary material B, which has two solutions labeled by $%
\sigma =\pm $:
\begin{equation}
n_{\sigma }^{2}=\varepsilon _{0}-\frac{g_{x}^{2}+g_{y}^{2}\mp \sqrt{%
(g_{x}^{2}+g_{y}^{2})^{2}+4\varepsilon _{0}^{2}g_{z}^{2}}}{2\varepsilon _{0}}%
.  \label{Qsolution}
\end{equation}

Let us analyze the solution (\ref{Qsolution}) for the different
polarization of the EMW and consider two limiting cases. First we
consider $\mathbf{g}=(0,0,g)$. Then
\begin{equation}
n_{\sigma }^{2}=\varepsilon _{0}\pm g,\qquad \mathbf{p}_{\sigma }=%
\begin{pmatrix}
\pm i \\
1 \\
0 \\
\end{pmatrix}%
.  \label{eqnc}
\end{equation}%
That solution formally describes two circularly polarized plane waves.

If $\mathbf{g}=(g_x,g_y,0)$ then
\begin{equation}  \label{eqnl}
n^2_+=\varepsilon_0,\qquad n^2_-=\varepsilon_0-\frac{g_x^2+g_y^2}{%
\varepsilon_0},
\end{equation}
and
\begin{equation}  \label{eqpl}
\mathbf{p}_+\propto%
\begin{pmatrix}
g_x \\
g_y \\
0 \\
\end{pmatrix}%
, \qquad \mathbf{p}_-\propto%
\begin{pmatrix}
i g_y \varepsilon_0 \\
- i g_x \varepsilon_0 \\
g_x^2+g_y^2 \\
\end{pmatrix}%
.
\end{equation}

Now we investigate reflection of EMW from the WS interface
when the $\mathbf{k}$ vector is perpendicular to the interface. WS occupies $%
z>0$ half-space, while the left-half space $z<0$ is an isotropic
dielectric with susceptibility $\varepsilon _{d}$. Then the electric
field{\ is given by}
\begin{eqnarray}
\mathbf{E} &=&\mathbf{p}_{d}\left( e^{i\omega n_{d}z/c-i\omega
t}-re^{-i\omega n_{d}z/c-i\omega t}\right) ,\quad z<0,  \notag  \label{eq:E}
\\
\mathbf{E} &=&A\mathbf{p}e^{i\omega nz/c-i\omega t},\qquad z>0,
\end{eqnarray}%
where $n_{d}=\sqrt{\varepsilon _{d}}$, $\mathbf{p}$ and $n$ here is either $%
\mathbf{p}_{+}$ ($n_{+}$) or $\mathbf{p}_{-}$ ($n_{-}$) and $r$ is the
reflection coefficient.

The magnetic field of this EMW {can be presented as}
\begin{eqnarray}
\mathbf{B} &=&\mathbf{q}_{d}\times \mathbf{p}_{d}\left( e^{i\omega
n_{d}z/c-i\omega t}+re^{-i\omega n_{d}z/c-i\omega t}\right) ,z<0,  \notag
\label{eq:H} \\
\mathbf{B} &=&A\mathbf{q}\times \mathbf{p}e^{i\omega nz/c-i\omega t},\qquad
z>0,
\end{eqnarray}%
where $\mathbf{q}_{d}=(0,0,n_{d}/c)$ and $\mathbf{q}=(0,0,n/c)$ and for
simplicity we again omit the index $\sigma =\pm $).

One can choose an arbitrary polarization vector of the incident wave
in the dielectric, $\mathbf{p}_d$, perpendicular to z-direction. But
we choose the specific $\mathbf{p}_d$ --- collinear to the
transverse components of polarization vector $\mathbf{p}$ in WS:
\begin{equation}  \label{eq:p}
\mathbf{p}_d =(p_x,p_y,0)/\sqrt{p_x^2+p_y^2},
\end{equation}
so $\mathbf{p}_d\times \mathbf{p}=0$.

Transverse components of electric field and magnetic field are continuous at
the interface. Then from Eqs.~\eqref{eq:E}-%
\eqref{eq:H} it follows,
\begin{equation}
1-r=A\sqrt{p_{x}^{2}+p_{y}^{2}},\quad n_{d}(1+r)=nA\sqrt{p_{x}^{2}+p_{y}^{2}}%
,
\end{equation}%
{and we get the reflection coefficient
\begin{equation}
r_{\sigma }=\frac{n_{d}-n_{\sigma }}{n_{d}+n_{\sigma }}.  \label{eq:r}
\end{equation}%
Here we restored the $\sigma $-index. One should pay attention that the
simple expression \eqref{eq:r} for} $r_{\sigma }$ is correct only for very
specific choice of incident wave polarization in the dielectric half-space.

First we consider the case when the gyration vector (or
$\mathbf{b}$) is parallel to the $\mathbf{k}$ vector of{\ the} EMW.
Then only circular polarized EMW with polarization $\mathbf{p}%
=(i,1,0)$ can propagate through WS. As it follows from Eq.~\eqref{eqnc}, $%
n_{-}^{2}<0$ for $g=g_{z}>\varepsilon _{0}^{\prime }$. Taking the
circular polarized wave with polarization $\mathbf{p}=(-i,1,0)$ as
the incident wave on WS surface we get{\ from Eq.~\eqref{eq:r},}
$|r_{-}|=1$ which means full reflection. The wave with polarization
$\mathbf{p}=(i,1,0)$ have finite probability of transmission through
WS interface, see Fig.~\ref{fig_r_bz}. So WS can serve as the
polarization filter that filters from the incident electromagnetic
irradiation circular polarized EMWs with certain polarization. The
direction of the electric field of a circular polarized light forms
a spiral in space. If the spiral twist in the direction of
$\mathbf{g}$ satisfies \textquotedblleft the right hand
rule\textquotedblright\ (clockwise) then this radiation can go
through WS, as it is shown in Fig.~\ref{figsp}\textit{a}.

Certain chirality $\pm 1$ can be attributed to circular polarization
depending on its ``right-hand'' or ``left-hand'' nature (wether the helix
describes the thread of a right-hand or left-hand screw, respectively). So
WS --- the host of chiral electrons is also the system with chiral
selectively of electromagnetic field, see Fig.~\ref{figsp}\textit{a} and
\textit{b}.

Now we consider the case when $\mathbf{b }\perp\mathbf{k}$. Then
only linear polarized EMW with
$\mathbf{p}=\mathbf{p}_+=(b_x,b_y,0)$, see Eq.~\eqref{eqpl}, can
propagate through WS. The wave with the orthogonal
polarization $\mathbf{p}_-=(-b_x,b_y,0)$ fades out if $\varepsilon_0<\sqrt{%
g_x^2+g_y^2}$, see Eq.~\eqref{eqnl}. Thus, in this case WS can serve
as the polarization filter that filters from the incident
electromagnetic irradiation linearly polarized EMWs along
$\mathbf{b}$.

In the intermediate case, when there is {a} certain nonzero angle between $%
\mathbf{b}$ and $\mathbf{k}$ (different from $\pi /2$), WS filters certain
elliptic polarized waves.

\textbf{Conclusions.} In this Letter we have obtained the frequency
dependencies of the refraction and reflection coefficients for the
Weyl semimetal for the different orientations of the vector
$\mathbf{b}$, which denotes the separation between the Weyl nodes of
opposite chirality in the first Brillouin zone and defines the
gyration vector $\mathbf{g}$. It is predicted that the propagation
of the normally incident polarized electromagnetic wave through the
WS for each polarization strongly depends on the orientation of the
polarization with respect to the gyration vector $\mathbf{g}$: when
the normally incident electromagnetic wave is collinear to the
gyration vector it is transmitted through the WS for one circular
polarization, while it is reflected for the opposite circular
polarization; the normally incident electromagnetic wave with the
linear polarization is transmitted through the WS if the
polarization is collinear to the gyration vector and reflected if
the polarization is normal to the gyration vector. The results of
calculations demonstrated that via changing the polarization of the
normally incident electromagnetic wave with respect to the vector
$\mathbf{b}$ the WS changes its behavior from the isotropic
dielectric to the medium with Kerr- or Faraday-like rotation of
polarization and also to the system with chiral selective
electromagnetic field. We propose that WSs can be applied as the
polarization filters.

%%%%%%%%%%%%%%%%%%%%%%%%%%%%%%%%%%%%%%%%%%%%%%%%%%%%%%%%%%%%%%%%%%%%%%%%%%%%%%%%%%%%%%%%%%%%%%%%%%%%%%%%%%%%%%%%%%%%%%%%%%%%%%%%%%%%%%%%%%%%%%%%%%%%%%%%%%%
%%%%%%%%%%%%%%%%%%%%%%%%%%%%%%%%%%%%%%%%%%%%%%%%%%%%%%%%%%%%%%%%%%%%%%%%%%%%%%%%%%%%%%%%%%%%%%%%%%%%%%%%%%%%%%%%%%%%%%%%%%%%%%%%%%%%%%%%%%%%%%%%%%%%%%%%%%%

\textbf{Acknowledgments.} N.M.C. was supported by the Russian
Science Foundation (grant 18-12-00438). O.L.B. and R.Ya.K. supported
by DOD  and PSC CUNY (grant W911NF1810433 and  grant 68660-00 46).
Yu.E.L. was supported by the Russian Foundation for Basic Research
(grant 20-02-00410).

%%%%%%%%%%%%%%%%%%%%%%%%%%%%%%%%%%%%%%%%%%%%%%%%%%%%%%%%%%%%%%%%%%%%%%%%%%%%%%%%%%%%%%%%%%%%%%%%%%%%%
%%%%%%%%%%%%%%%%%%%%%%%%%%%%%%%%%%%%%%%%%%%%%%%%%%%%%%%%%%%%%%%%%%%%%%%%%%%%%%%%%%%%%%%%%%%%%%%%%%%%%

%\bibliographystyle{pss}
%\bibliography{OLEGbib}

\begin{thebibliography}{99}
\bibitem{Liu_14} Z. K. Liu, B. Zhou, Y. Zhang, Z.~J. Wang, H.~M. Weng, D.
Prabhakaran, S.~K. Mo, Z.~X. Shen, Z. Fang, X. Dai, Z. Hussain, and Y.~L.
Chen, Science \textbf{343}, 864 (2014).

\bibitem{Borisenko} S. Borisenko, Q. Gibson, D. Evtushinsky, V. Zabolotnyy,
B.~B\"{u}chner, and R.~J. Cava, Phys. Rev. Lett. \textbf{113}, 027603 (2014).

\bibitem{Neupane} M. Neupane, S.~Y. Xu, R. Sankar, N. Alidoust, G. Bian, C.
Liu, I. Belopolski, T. R. Chang, H. T. Jeng, H. Lin, A. Bansil, F. Chou, and
M. Z. Hasan, Nat. Commun. \textbf{5}, 3786 (2014).

\bibitem{Xiong}
J. Xiong, S. K. Kushwaha, T. Liang, J. W. Krizan, M. Hirschberger, W. Wang, R. J. Cava, N. P. Ong, Science \textbf{350}, 413 (2015).

\bibitem{Wang}
L. Wang, C. Li, D. Yu, et al., Nat. Commun. \textbf{7}, 10769 (2016).

\bibitem{Liang} T. Liang, J. Lin, Q. Gibson, T. Gao, M. Hirschberger, M. Liu, R.J. Cava, and N.P.
Ong,
Phys. Rev. Lett. \textbf{118}, 136601 (2017).

\bibitem{Galletti}
L. Galletti, T. Schumann, O. F. Shoron, M. Goyal, D. A. Kealhofer, H. Kim, and S. Stemmer,
Phys. Rev. B \textbf{97}, 115132 (2018).

\bibitem{Zhang} S.S.-L. Zhang, A. A. Burkov, I. Martin, and O. G.
Heinonen,
Phys. Rev. Lett. \textbf{123}, 187201 (2019).

\bibitem{Puphal} P. Puphal, V. Pomjakushin, N. Kanazawa, V. Ukleev, D. J. Gawryluk, J. Ma, M. Naamneh, N. C. Plumb, L. Keller, R. Cubitt, E. Pomjakushina, and J. S.
White, Phys. Rev. Lett. \textbf{124}, 017202 (2020).

\bibitem{Wan} X. Wan, A.~M. Turner, A. Vishwanath, and S.~Y. Savrasov, Phys.
Rev. B \textbf{83}, 205101 (2011).

%\bibitem{Kane} C.~L. Kane and E.~J. Mele, Phys. Rev. Lett. \textbf{95}, 226801 (2005).

\bibitem{Young} S.~M. Young, S. Zaheer, J.~C.~Y. Teo, C.~L. Kane, E.~J.
Mele, and A.~M. Rappe, Phys. Rev. Lett. \textbf{108}, 140405 (2012).


\bibitem{Abanov} G.~M. Monteiro, A.~G. Abanov, and D.~E. Kharzeev, Phys.~Rev.~B {\bf 92}, 165109 (2015).

\bibitem{Polini} F.~M.~D. Pellegrino, M.~I. Katsnelson, and M. Polini, Phys.
Rev. B \textbf{92}, 201407 (2015).

\bibitem{Kotov} O.~V. Kotov and Yu.~E. Lozovik, Phys.~Rev. B {\bf 98}, 195446 (2018).


\bibitem{Belyanin} Q. Chen, A.~R. Kutayiah, I. Oladyshkin, M. Tokman, and A.
Belyanin, Phys. Rev. B \textbf{99}, 075137 (2019).

\bibitem{Folkes} B. Cheng, P. Taylor, P. Folkes, C. Rong, and N.~P.
Armitage, Phys. Rev. Lett. \textbf{122}, 097401 (2019).

\bibitem{Yan} B. Yan and C. Felser, Annu. Rev. Condens. Matter Phys. \textbf{%
8},337 (2017).

\bibitem{Armitage_rmp} N.~P. Armitage, E.~J. Mele, and A.
Vishwanath, Rev.~Mod.~Phys. {\bf 90}, 015001 (2018).


\bibitem{Belyakov} V.~A. Belyakov, V.~E. Dmitrienko, and V.~P. Orlov, Soviet
Physics Uspekhi \textbf{22}, 64 (1979).

\bibitem{Vetrov} S.~Ya. Vetrov, I.~V. Timofeev, and V.~F. Shabanov, Physics Uspekhi \textbf{63}, 1 (2020).

\bibitem{Vafek} O. Vafek and A. Vishwanath,  Annu. Rev. Condens. Matter Phys. \textbf{5}, 83 (2014).

\bibitem{Wehling} T. Wehling, A. Black-Schaffer, and A. Balatsky, Adv. Phys.
\textbf{63}, 1 (2014).

\bibitem{Nielsen1} H. Nielsen and M. Ninomiya, Nucl. Phys. B \textbf{185}, 20 (1981).

\bibitem{Nielsen2} H. Nielsen and M. Ninomiya, Nucl. Phys. B \textbf{193}, 173 (1981).

\bibitem{Nielsen3} H. Nielsen and M. Ninomiya, Phys. Lett. B \textbf{130}, 389 (1983).

\bibitem{japan} K. Fujikawa, Phys. Rev. Lett. \textbf{42}, 1195 (1979).

\bibitem{Burkov} A. A. Zyuzin and A. A. Burkov, Phys. Rev. B \textbf{86},
115133 (2012).

\bibitem{Tewari} P. Goswami and S. Tewari, Phys. Rev. B \textbf{88}, 245107
(2013).

\bibitem{Hosur} P. Hosur and X. Qi, C. R. Phys. \textbf{14}, 857
(2013).

\bibitem{Wilczek} F. Wilczek, Phys. Rev. Lett. \textbf{58}, 1799
(1987).

\bibitem{Grushin} A.~G. Grushin, Phys. Rev. D \textbf{86}, 045001 (2012).

\bibitem{Hosur2} P. Hosur and X.~L. Qi, Phys. Rev. B \textbf{91}, 081106
(2015).

\bibitem{Trivedi} M. Kargarian, M. Randeria, and N. Trivedi, Sci. Rep.
\textbf{5}, 12683 (2015).

\bibitem{Zyuzin} A.~A. Zyuzin and V.~A. Zyuzin, Phys. Rev. B \textbf{92},
115310 (2015).



\bibitem{Sharma} P. Goswami, G. Sharma, and S. Tewari, Phys. Rev. B \textbf{%
92}, 161110 (2015).

\bibitem{DasSarma} J. Hofmann and S. Das Sarma, Phys. Rev. B \textbf{93},
241402 (2016).

\bibitem{Rosenstein} B. Rosenstein and M. Lewkowicz, Phys. Rev. B \textbf{88}%
, 045108 (2013).

\bibitem{landau2013electrodynamics} L.~D. Landau, J. Bell, M. Kearsley, L.
Pitaevskii, E. Lifshitz, and J. Sykes, Electrodynamics of continuous media
(Elsevier, 2013).
\end{thebibliography}

\end{document}